\documentclass[aps,prd,twocolumn,showpacs]{revtex4}
\usepackage[T1]{fontenc}
\usepackage{ae,aecompl}

\usepackage{amsmath}
\usepackage{graphicx}	
\usepackage{lscape}
\usepackage{indentfirst}

\usepackage{amssymb}
\usepackage{bm}
\usepackage{xcolor}

\newcommand{\be}{\begin {equation}}
\newcommand{\ee}{\end {equation}}
\newcommand{\beq}{\begin {eqnarray}}
\newcommand{\eeq}{\end {eqnarray}}
\newcommand{\ff}{f} 
\newcommand{\ii}{i} 

\newcommand{\apss}{Astrophys.\ Space Sci.}
\newcommand{\aap}{Astron.\ Astrophys.}
\newcommand{\mnras}{Mon.\ Not.\ R.\ Astron.\ Soc.}
\newcommand{\apjs}{Astrophys.\ J.\ Suppl.\ Ser.}

\newcommand{\apjl}{Astrophys.\ J.}
\newcommand{\aapr}{Astron.\ Astrophys.\ Rev.}
\newcommand{\apspr}{Astrophys.\ Space Phys.\ Res.}
\newcommand{\physrep}{Phys.\ Rep.}
\newcommand{\ssr}{Space Sci.\ Rev.}

\renewcommand{\d}{d} 

\begin{document}

\title{Statistical features of multiple Compton scattering in a strong magnetic field}

\author{A.~A.~Mushtukov$^{1,2}$}
\email{alexander.mushtukov@physics.ox.ac.uk} 
\author{I.~D.~Markozov$^{3}$}  
\author{V.~F.~Suleimanov$^{4}$}
\author{D.~I.~Nagirner$^{3}$}
\author{A.~D.~Kaminker$^{5}$}
\author{A.~Y.~Potekhin$^{5}$}
\author{S.~Portegies Zwart$^{2}$} 

\affiliation{
$^1$ Astrophysics, Department of Physics, University of Oxford, Denys Wilkinson Building, Keble Road, Oxford OX1 3RH, UK\\
$^2$ Leiden Observatory, Leiden University, NL-2300RA Leiden, The Netherlands \\
$^3$ Sobolev Astronomical Institute, Saint Petersburg State University, Saint-Petersburg 198504, Russia \\      
$^4$ Institut f\"ur Astronomie und Astrophysik, Universit\"at T\"ubingen, 
    Sand 1, D-72076 T\"ubingen, Germany \\
$^5$ Ioffe Institute, Politekhnicheskaya 26, St Petersburg 194021, Russia
} 

\date{\today}

\begin{abstract}
Compton scattering is a key process shaping spectra formation and accretion flow dynamics in accreting strongly magnetized neutron stars (NSs).
A strong magnetic field affects the scattering cross section and makes it dependent on photon energy, momentum, and polarization state.
Using Monte Carlo simulations, we investigate statistical features of Compton scattering of polarized X-ray radiation in a strong magnetic field.
Our analysis is focused on photon gas behaviour well inside the scattering region.
We take into account the resonant scattering at the fundamental cyclotron frequency, thermal distribution of electrons at the ground Landau level, and bulk velocity of the electron gas.
We show that 
(i) the photons scattered around the cyclotron energy by the electron gas at rest tend to acquire the final energy close to the cyclotron one with very small dispersion measure;
(ii) the redistribution of photons within the Doppler core of cyclotron resonance differs significantly from the complete redistribution; 
(iii) the efficiency of momentum transfer from photons to the electron gas is affected by the temperature of electron gas both for photons at cyclotron energy and below it;
(iv) the momentum transfer from photons to the electron gas of non-zero bulk velocity is more efficient in the case of magnetic scattering. 
\end{abstract}


\maketitle



\section{Introduction}

Compton scattering of X-ray photons is the primary process shaping interaction of radiation and matter 
at high temperatures in many astrophysical objects \citep{1979rpa..book.....R}.
Magnetic Compton scattering \citep{1979PhRvD..19.2868H,1986ApJ...309..362D,2016PhRvD..93j5003M} is a key process standing behind spectra and polarization formation in isolated and accreting strongly magnetized neutron stars (NSs) -- X-ray pulsars (XRPs; see, e.g., \citep{1992ans..book.....L,2015A&ARv..23....2W,2022arXiv220414185M}).
A strong magnetic field $\bm{B}$ modifies the elementary processes
on the level of quantum wave functions (see \citep{1992herm.book.....M,2006RPPh...69.2631H} for review) and changes dramatically their cross sections.
In particular, the Compton scattering cross section becomes strongly dependent on the magnetic field strength $B$, photon energy, direction of the photon momentum $\bm{k}$ with respect to $\bm{B}$, and polarization state \citep{1971PhRvD...3.2303C,1979PhRvD..19.2868H,1986ApJ...309..362D,2016PhRvD..93j5003M}.
The electrons occupy the Landau levels, and the electron transitions between the levels lead to resonances in the scattering events.
The scattering cross section at resonant energies exceeds the Thomson cross section $\sigma_{T}$ by orders of magnitude \citep{1991ApJ...374..687H}.

The resonant Compton scattering results in the appearance of cyclotron scattering features in the energy spectra of magnetized NSs \citep{2019A&A...622A..61S} and affects the interaction between accreting material and radiation in 
the XRPs \citep{2015MNRAS.447.1847M}. 
Because the $B$-field strength determines the energies where the resonances appear, the detection of cyclotron scattering features in X-ray spectra is widely used to probe the magnetic field strength of NSs.

The elementary scattering was considered in detail both in
the non-relativistic limit \citep{1971PhRvD...3.2303C,1979PhRvD..19.2868H} and taking the effects of special relativity into account \citep{1986ApJ...309..362D,2016PhRvD..93j5003M,2017A&A...597A...3S,2017A&A...601A..99S}.
Harding and Daugherty \citep{1991ApJ...374..687H} compared QED polarization averaged cross sections of Compton scattering and cyclotron absorption. 
Useful analytical approximations were obtained in
a particular case of scattering of photons moving along the field lines \citep{2000ApJ...540..907G}.
{The redistribution of X-ray photons in the cyclotron lines was discussed by Wasserman and Salpeter \citep{1980ApJ...241.1107W}, neglecting polarization and assuming a constant temperature.
They  obtained  analytical approximations for angle-averaged radiation  field in close proximity to the cyclotron resonance.}

In general, because of the complicated behavior of Compton scattering in a strong magnetic field, models of radiative transfer 
with allowance for magnetic Compton scattering are still limited, especially as concerns multiple scattering.

{Some progress in solving problems of radiation transfer in an optically thick highly magnetized plasma has been achieved using an approximation of coupled diffusion of  normal  modes (see,  e.g.   Nagel  \citep{1980ApJ...236..904N,1981ApJ...251..278N}).   Kaminker et~al.\  \citep{1982Ap&SS..86..249K,1983Ap&SS..91..167K} also  applied  this  approach in the ``cold plasma approximation'', i.e., considering  the coherent scattering of photon  energies outside the Doppler cores  of  the  cyclotron resonances. They considered a wide range of the energies  taking into account  the effects of electron-positron vacuum polarization in a strong magnetic field (e.g.  Adler \citep{1971AnPhy..67..599A},  Meszaros  and  Ventura  \citep{1978PhRvL..41.1544M},  Gnedin et~al.\ \citep{1978SvAL....4..117G} ).}


{Transport of polarized X-ray radiation in a hot, highly magnetized plasma at energies well  below  the cyclotron energy in the diffusion approximation was investigated by  Lyubarskii \citep{1986Ap.....25..577L} with  taking into account multiple Compton scattering and assuming linearly polarized normal modes.}


Monte Carlo simulations of polarized (in two polarization modes) radiative transfer in cylindrical geometry were initiated in late 70s by Yahel \citep{1979ApJ...229L..73Y,1980ApJ...236..911Y} under assumptions of fixed temperature, zero velocity, and the Planck spectrum as a source of seed photons.
Later Monte Carlo approach was applied in models of cyclotron scattering feature formation in spectra of XRPs \citep{1999ApJ...517..334A} and in models describing spectra of XRPs at extremely low mass accretion rates \citep{2021MNRAS.503.5193M}.
Monte Carlo simulations of resonant Comptonization were applied to explain non-thermal X-ray emission of magnetars \citep{2007ApJ...660..615F,2008MNRAS.386.1527N,2011ApJ...730..131F,2014MNRAS.438.1686T}, where the X-ray energy spectrum in the quiescent state is thought to arise from resonant Compton scattering of thermal photons by charges moving in a twisted magnetosphere of a compact object.

On the base of the Feautrier numerical scheme \citep{1978stat.book.....M}, Nagel \citep{1981ApJ...251..278N,1981ApJ...251..288N} studied radiative transfer in  two specific geometries: the slab and the cylinder. 
The calculations were performed either
taking into account angular redistribution of photons but 
assuming coherent scattering \citep{1981ApJ...251..278N},
or with allowance for the energy exchange in scattering events but neglecting angular redistribution \citep{1981ApJ...251..288N}.
These calculations were improved later by M\'esz\'aros and Nagel \citep{1985ApJ...298..147M,1985ApJ...299..138M}, who have taken into account both energy exchange and angular redistribution of X-ray photons due to the scattering,
as well as the vacuum polarization effects. 
They derived some predictions for the X-ray spectrum and its variations with the direction.
It was shown that the effects of vacuum polarization influence both polarization and depth of the cyclotron scattering feature.

{The radiation transfer in an optically thick hot plasma, in which incoherent Compton scattering plays an important role, was investigated by Pavlov et~al.\cite{1989PhR...182..187P}. They showed, in particular,  that in a strong magnetic field, the process of Comptonization significantly affects the characteristics of radiation, mainly in the vicinity of the cyclotron resonance.
It should be noted, however, that all authors cited above 
solved the radiative transfer problem under assumptions of constant temperature, constant mass density in the atmosphere, non-relativistic Maxwellian distribution of electrons, and the electron gas being at rest.}

Recently the Feautrier numerical scheme was improved further and succesfully applied for calculations of polarized spectra of XRPs at low luminosity states \citep{2021A&A...651A..12S}.

Alexander et al. \citep{1989ApJ...342..928A} analyzed the influence of induced scattering on the spectra of magnetized atmospheres. 
Both non-relativistic and relativistic cross sections were used and compared in the paper.

Burnard et al. \citep{1990ApJ...349..262B} solved the radiative transfer problem with allowance for magnetic Comptonization, temperature variations in the atmosphere, and arbitrary
inclination of the magnetic field to the atmosphere.
The effects of vacuum polarization were not taken into account. 
The authors developed a novel moment-Feautrier method for solution of the radiative transfer problem.
They obtained broadband spectra for various orientations of the $B$-field and calculated the dependence
of temperature and mass density on the optical depth.

The relativistic kinetic equation for magnetized Compton scattering with the proper account of the induced processes and Pauli exclusion principle was proposed by Mushtukov et al. \citep{2012PhRvD..85j3002M}, but never used in numerical calculations because of its complexity. 

In this paper, we explore statistical properties of multiple Compton scattering in a strong magnetic field using Monte Carlo simulations.
We investigate how multiple scattering events in a medium with given physical parameters influence the distribution of X-ray photons 
over the energy and momentum.
We do not consider photon's escape from the medium, and in that sense our analysis is focused on photons behaviour well inside the scattering region. 
However, the conclusions on the statistical features of a single scattering event can be applicable to the cases of optically thin medium.
We examine features of both resonant and non-resonant scattering,
taking into account polarization of X-ray photons, thermal motion of electrons along $B$-field lines, and bulk motions of material, which is particularly important for spectra formation and accretion flow dynamics in the XRPs.
Our simulations allow to reveal specific effects of photon redistribution due to the scattering in a strong magnetic field and test applicability of some assumptions used in the radiative transfer theory (in particular, the assumption of complete redistribution within the Doppler core of a line).
Our analysis is limited to non-relativistic scattering cross sections and is performed under the
assumption that the majority of electrons occupy the ground Landau level.
This allows us to use simplified scattering cross sections with resonances at the cyclotron fundamental only.
The assumption of predominant occupation of the ground Landau level is valid in a strong $B$-field, because the radiative de-excitation occurs much faster than the collisional excitation \citep{1979A&A....78...53B,1992herm.book.....M}.
The used assumptions are adequate for studying photon redistribution due to the scattering in the atmospheres and accretion columns of the most typical strongly magnetized neutron stars.

The paper is composed as follows. In Sec.~\ref{sec:PolModes} we recall the formalism of normal electromagnetic modes in magnetized plasmas. 
In Sec.~\ref{sec:CrossSection} we derive working formulas for cross sections of photon scattering by an ensemble of (in general, moving and relativistic) electrons. 
Section~\ref{sec:RedistribMomenta} presents expressions for the total cross section and statistical moments of the distribution of scattered photons. 
Our Monte Carlo code is briefly described in Sec.~\ref{sec:MonteCarlo}. 
Numerical results are presented and discussed in Sec.~\ref{sec:NumRes}. 
Potential astrophysical applications of the results
are listed in Sec.~\ref{sec:AstroApp}. The summary is given in Sec.~\ref{sec:Summary}. Appendixes A and B give analytic expressions, respectively, for the electron momentum distribution and for the scattering amplitudes, which are used in our work.

\section{Polarization modes of X-ray photon in a strongly magnetized plasma}
\label{sec:PolModes}

The photon in a strongly magnetized plasma is described by its polarization state $s$, energy $k$, and direction of momentum given by two angles: the angle between the $B$-field and photon momentum $\theta$ and the azimuthal angle $\varphi$.
Thus, the photon momentum in the Cartesian coordinates, where the $z$ axis is aligned with the direction of magnetic field is given by 
\beq
\bm{k}=k(\sin\theta\cos\varphi,\sin\theta\sin\varphi,\cos\theta).
\eeq
Here and hereafter, unless otherwise stated, we use the natural system of units with the speed of light, Planck and Boltzmann constants, and electron mass equal to unity  ($c=\hbar=k_{\rm B}=m_e=1$),
and neglect the difference between the phase and group velocities.

Plasma in a strong magnetic field is anisotropic and birefringent (see reviews \cite{2006RPPh...69.2631H,1984ASPRv...3..197P}). 
The propagation of photons is determined by the dielectric tensor and the
magnetic permeability tensor, which are affected by specific properties of magnetized plasma and magnetized vacuum.
These tensors are determined by plasma mass density $\rho$, chemical composition, magnetic field $\bm{B}$, and temperature $T$.
For a sector in the parameter space $\{\rho,B,T\}$, covering the most typical conditions in the outer layers of the strongly magnetized NSs,
the electromagnetic waves propagate in the form of two orthogonal normal modes of different phase and group velocities: 
the ordinary mode (O-mode)
and 
the extraordinary mode (X-mode) 
(see, e.g., \cite{1974JETP...38..903G,1992herm.book.....M}).

To describe polarization of a photon, let us use the Cartesian coordinate system, where the $z$ axis is aligned with the direction of photon momentum $\bm{k}$, the $x$ axis is taken to be perpendicular to the $\bm{k}-\bm{B}$ plane, the $y$ axis belongs 
to the $\bm{k}-\bm{B}$ plane and complements the coordinate system to the right-hand one. 
The electric vector of an elliptically polarized electromagnetic wave can be represented locally as 
\beq\label{eq:Pol_Representation}
\bm{E}=(E_{x}\mathbf{e}_{x}+E_{y}\mathbf{e}_{y})e^{-i\omega t}, 
\eeq
where $\mathbf{e}_{x},\mathbf{e}_{y}$ are unit vectors along the $x$ and $y$ axis respectively, $E_x,E_y\in \mathbb{C}$, and $\omega$ is the photon frequency.

\begin{figure}
\centering 
\includegraphics[width=\columnwidth]{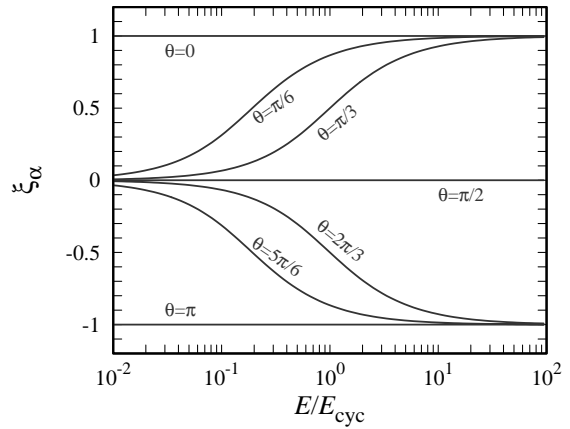} 	
\caption{The ellipticity parameter of the X-mode ($\alpha=1$, see equation \ref{eq:NormWavEll}) as a function of photon energy $E$. Different curves are given for different angles $\theta$ between the photon momentum and $B$-field direction.}
\label{pic:sc_NormWavEll_01}
\end{figure}

In the case of pure vacuum in a strong magnetic field, the normal modes are linearly polarized. 
The electric field vector of the X-mode photons oscillates in the direction perpendicular to the $\bm{k}-\bm{B}$ plane (i.e., $E_{y}=0$), 
while the electric field vector of the O-mode oscillates within the $\bm{k}-\bm{B}$ plane (i.e., $E_{x}=0$).
In general, the modes are elliptically polarized.
Their ellipticity depends on $\bm{k}$ and $\bm{B}$,
and on the plasma density.
In the case of dominating plasma effects and neglecting the effects of vacuum polarization,
the ellipticity parameter, establishing the relation between $E_x$ and $E_y$ in (\ref{eq:Pol_Representation}), can be approximated as
\beq\label{eq:NormWavEll}
\xi_\alpha(E,\theta) &\equiv& -i\left(\frac{E_y}{E_x}\right)_{\alpha}  \\ 
&=& \frac{2\cos\theta}{\frac{E_\mathrm{cyc}}{E}\sin^2\theta -(-1)^{\alpha} \sqrt{\frac{E^2_\mathrm{cyc}}{E^2}\sin^4\theta +4\cos^2\theta} }, \nonumber
\eeq
where $\alpha=1$ for X-mode and $\alpha=2$ for O-mode in plasma, $E_\mathrm{cyc}\approx 11.6\,(B/10^{12}\,\mbox{G})$~keV (see, e.g., \cite{1974JETP...38...51G} and see Fig.\,\ref{pic:sc_NormWavEll_01}).
Note that $\xi_1=-\xi_2^{-1}$.

The electric field vectors of elliptically polarized plasma modes $\bm{E}_{1,2}^{(p)}$ can be expressed through the electric field vectors of linearly polarized modes $\bm{E}_{1,2}^{(v)}$ as
\beq
\bm{E}_{2}^{(p)}=\frac{ i\bm{E}_{1}^{(v)} + \xi_{2} \bm{E}_{2}^{(v)} }{(1+|\xi_{2}|^2)^{1/2}},\quad
\bm{E}_{1}^{(p)}=\frac{ -i \xi_{2}\bm{E}_{1}^{(v)} + \bm{E}_{2}^{(v)} }{(1+|\xi_{2}|^2)^{1/2}},
\eeq
which can be rewritten as
\beq
\left(\begin{array}{c} \bm{E}_{1}^{(p)} \\ \bm{E}_{2}^{(p)} \end {array}\right) 
= \widehat{M}_\mathrm{pv} \left(\begin{array}{c} \bm{E}_{1}^{(v)} \\ \bm{E}_{2}^{(v)} \end {array}\right) ,
\eeq
where
\beq\label{eq:Transform_Matrix}
\widehat{M}_\mathrm{pv} &=& 
\frac{1}{\sqrt{1+|\xi_{2}|^2}}\left(\begin {array}{cc} -i \xi_{2} & 1 \\ i & \xi_{2} \end {array} \right) \nonumber \\
 &=& 
\frac{{\rm sign}(\xi_{1})}{\sqrt{1+|\xi_{1}|^2}} \left(\begin {array}{cc}  i &  \xi_{1} \\ i  \xi_{1} & -1 \end {array} \right).
\eeq
is unitary matrix (i.e., $\widehat{M}_\mathrm{pv}^{-1}=\widehat{M}_\mathrm{pv}^{\dagger}$).
The linearly polarized modes are chosen here to have the electric vector perpendicular ($\bm{E}_{1}^{(v)}$) or belonging ($\bm{E}_{2}^{(v)}$) to the $\bm{k}-\bm{B}$ plane.

\begin{figure}
\centering 
\includegraphics[width=\columnwidth]{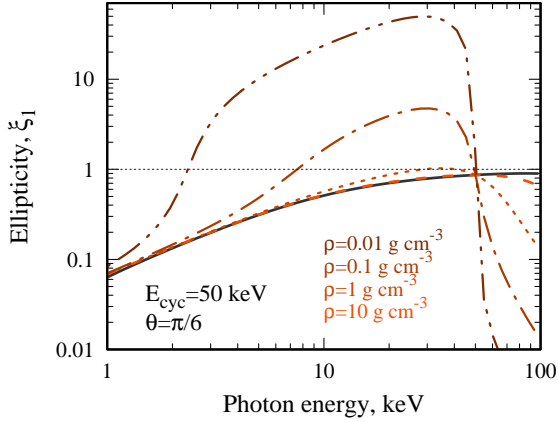} 	
\caption{
Comparison between the approximate ellipticity of photon with $\alpha=1$, calculated with allowance for the plasma effect only (black solid line), and the ellipticity calculated  with allowance for both plasma and vacuum effects.
Different lines are given for different mass densities: $0.01$ (dashed-double-dotted), $0.1$ (dashed-dotted), $1$ (dotted), and $10$  g\,cm$^{-3}$ (dashed).
The expression (\ref{eq:NormWavEll}) provides a reasonable approximation for the ellipticity at relatively large mass densities. 
Parameters: fully ionised hydrogen plasma, $E_\mathrm{cyc}=50$~keV, $\theta=\pi/6$.
}
\label{pic:ellipt_compar}
\end{figure}

Expression (\ref{eq:NormWavEll}) provides a reasonable approximation for the ellipticity of normal modes in a sufficiently dense magnetized plasma. 
At low mass densities, the polarization of normal modes is strongly affected or even dominated by the  effects of vacuum polarization (see Fig.\,\ref{pic:ellipt_compar}, where the vacuum influence on the polarization state of normal modes was taken into account according to \citep{2006MNRAS.373.1495V}).
Below we study statistical features of Compton scattering neglecting effects of vacuum polarization and assuming that the ellipticity of normal modes is given by Eq.~(\ref{eq:NormWavEll}).

\section{Scattering cross section and the redistribution of X-ray photons}
\label{sec:CrossSection}

Each scattering event results in a change of photon energy, momentum, and polarization state.
Let us designate the conditions before and after scattering by subscripts ``i" (``initial") and ``f" (``final"), respectively.
There are two conserved physical quantities in magnetic Compton scattering: the total energy and the momentum along the direction of magnetic field.
The corresponding conservation laws can be written as
\beq\label{eq:ConsLaws}
E_{n_\ii}(p_{z,\ii})+k_\ii&=&E_{n_\ff}(p_{z,\ff})+k_\ff,\\
p_{z,\ii}+k_{z,\ii}&=&p_{z,\ff}+k_{z,\ff},
\nonumber
\eeq
where $E_n(p_z)=(1+p_z+2bn)^{1/2}$ is the electron energy at the Landau level $n$, 
$b$ is the field strength in the natural units (in the ordinary units, $b=B/B_{Q}$, where $B_{Q}\approx 4.414\times 10^{13}$~G is the critical QED field strength), 
$p_z$  is the  electron momentum along the magnetic field, and $k_{z}=k\cos\theta$ is the $z$-component of $\bm{k}$. 

The redistribution of photons over momentum and polarization states due to the scattering is described by the double differential cross section:
\beq
\label{eq:DDiffCS00}
\frac{\d\sigma_{s_{f},s_{i}}}{\d\mathbf{\Omega}_{f}\d k_{f}}
(p_{z,i},k_{i},\mathbf{\Omega}_{i},k_{f},\mathbf{\Omega}_{f}),	
\eeq
where $\mathbf{\Omega}_{i,f}=\{\theta_{i,f},\varphi_{i,f}\}$ determines the direction of photon momentum, 
$s_{i,f}$ are photon polarization states. 

If a photon of momentum $\bm{k}_{i}$ is scattered by an electron of momentum $p_{z,i}$, 
there is a strict relation between the final photon energy $k_{f}$ and the direction of its momentum $\mathbf{\Omega}_{f}$, which can be obtained from Eq.~(\ref{eq:ConsLaws}).
In this particular case the double differential cross section is given by
\begin{align}
\label{eq:DiffCrossSection01} 
&\frac{\d\sigma_{s_{f},s_{i}}}{\d\mathbf{\Omega}_{f}\d k_{f}}
(p_{z,i},k_{i},\mathbf{\Omega}_{i},k_{f},\mathbf{\Omega}_{f})
= \\
&\quad\quad\quad=
\delta[k_{f}-k_{f}(p_{z,i},k_{i},\theta_{i},\theta_{f})]\,
\frac{\d\sigma_{s_{f},s_{i}}}{\d\mathbf{\Omega}_{f}}
(p_{z,i},k_{i},\mathbf{\Omega}_{i},\mathbf{\Omega}_{f}).
\nonumber
\end{align}
To obtain the total scattering cross section, one has to integrate both over the solid angle $\mathbf{\Omega}_{f}$ and final photon energy $k_{f}$.
Photons of a given momentum $\bm{k}_{i}$, being scattered into a direction $\mathbf{\Omega}_{f}$ by an ensemble of electrons with different $p_z$, acquire a specific energy distribution, which is determined by the distribution of electrons over $p_z$. 
In our simulation we assume that the electrons occupy only the ground Landau level and their distribution over $p_z$ in the reference frame co-moving with a gas is given by the one-dimensional Maxwell distribution (see Appendix\,\ref{App:Maxwell} and Fig.\,\ref{pic:Maxwell}), i.e., the distribution is determined by two parameters: temperature of the gas $T$ and its velocity $\beta$.
If the latter distribution is $f_{e}(p_z)$ ($\int_{-\infty}^{+\infty} f_{e}(p_z)\d p_z = 1$), the differential ({\it ensemble averaged}) cross section
describing the redistribution of photons over the directions  can be 
written as
\begin{align}
\label{eq:DiffCrossSection02}
&\frac{\d\sigma_{s_{f},s_{i}}}{\d\mathbf{\Omega}_{f}} ([f_{e}(p_z)],k_{i},\mathbf{\Omega}_{i},\mathbf{\Omega}_{f})
= \\
&\quad\quad\quad= \int\limits_{-\infty}^{+\infty}\d p_{z,i}\,f_{e}(p_{z,i})
\frac{\d\sigma_{s_{f},s_{i}}}{\d\mathbf{\Omega}_{f}} (p_{z,i},k_{i},\mathbf{\Omega}_{i},\mathbf{\Omega}_{f}).
\nonumber
\end{align}
On the other hand, the same cross section can be 
obtained by integration 
of  {\it ensemble averaged} 
Eq.~(\ref{eq:DiffCrossSection01}) over the final photon energy:
\begin{align}
\label{eq:DiffCrossSection03}
&\frac{\d\sigma_{s_{f},s_{i}}}{\d\mathbf{\Omega}_{f}} ([f_{e}(p_z)],k_{i},\mathbf{\Omega}_{i},\mathbf{\Omega}_{f})
=\\
&=\int\limits_{0}^{\infty}\d k_{f}\, 
\frac{\d\sigma_{s_{f},s_{i}}}{\d\mathbf{\Omega}_{f}\d k_{f}}([f_{e}(p_z)],k_{i},\mathbf{\Omega}_{i},k_{f},
\mathbf{\Omega}_{f})
= \nonumber \\
&=\int\limits_{-\infty}^{+\infty}\d p_{z,i}\,f_{e}(p_{z,i})
\int\limits_{0}^{\infty}\d k_{f}\
\frac{\d\sigma_{s_{f},s_{i}}}{\d\mathbf{\Omega}_{f}\d k_{f}}(p_{z,i},k_{i},\mathbf{\Omega}_{i},k_{f},\mathbf{\Omega}_{f}).  \nonumber
\end{align}
Using Eq.~(\ref{eq:DiffCrossSection01}) in the second integral
over $\d k_f$  we come to Eq.~(\ref{eq:DiffCrossSection02}). 
Based on 
Eqs.~(\ref{eq:DiffCrossSection02}) and  
(\ref{eq:DiffCrossSection03})   
we will get  in Sec.~\ref{sec:RedistribMomenta}
the original expressions for further calculations.

The Lorentz transformations give the relation between the
cross section of scattering by an electron with momentum $p_z$ and by an electron at rest
(the values marked with asterisks): 
\begin{align}
\label{eq:diff_cs_pz}
&\frac{\d\sigma_{s_{f},s_{i}}}{\d\mathbf{\Omega}_{f}}(p_{z,i},k_{i},\mathbf{\Omega}_{i},\mathbf{\Omega}_{f})
=\\
&\quad\quad= \frac{\d\sigma_{s_{f},s_{i}}}{\d\mathbf{\Omega}^*_{f}}(p_{z,i}=0,k^*_{i},\theta^*_{i},\varphi_{i},\theta^*_{f},\varphi_{f})
\frac{(1-\beta^2)}{(1-\beta\cos\theta_{f})^2},
\nonumber
\end{align}
where 
\beq 
k^*_{i}=k_{i}\gamma (1-\beta\cos\theta_{i}),\quad
\cos\theta^*_{i,f}=\frac{\cos\theta_{i,f}-\beta}{1-\beta\cos\theta_{i,f}}
\label{Doppler}
\eeq
determine the values of the photon momentum and cosine of colatitude, respectively, in the reference frame of an electron,
$\gamma=(1+p_z^2)^{1/2}$ is the electron energy, and $\beta=p_z/\gamma$ is the electron velocity
in the $z$ direction (in the natural units).

The differential cross section of Compton scattering in a strong magnetic field is obtained in quantum electrodynamics \citep{1971PhRvD...3.2303C,1979PhRvD..19.2868H,1986ApJ...309..362D,2000ApJ...540..907G,2016PhRvD..93j5003M,2017A&A...597A...3S,2017A&A...601A..99S},
and is related to the complex amplitudes of scattering $a_{s_{f}s_{i}}$:
\beq 
\label{eq:Amp2CS}
\frac{\d\sigma_{s_{f}s_{i}}}{\d\mathbf{\Omega}_{f}}(p_{z,i}=0,k_{i},\mathbf{\Omega}_\ii,\mathbf{\Omega}_\ff)
=\frac{3}{32\pi}\sigma_{T}|a_{s_{f}s_{i}}|^2.
\eeq
The amplitudes depend on the exact expression for the polarization modes (see Section\,\ref{sec:PolModes}).

\begin{figure*}
\centering 
\includegraphics[width=\textwidth]{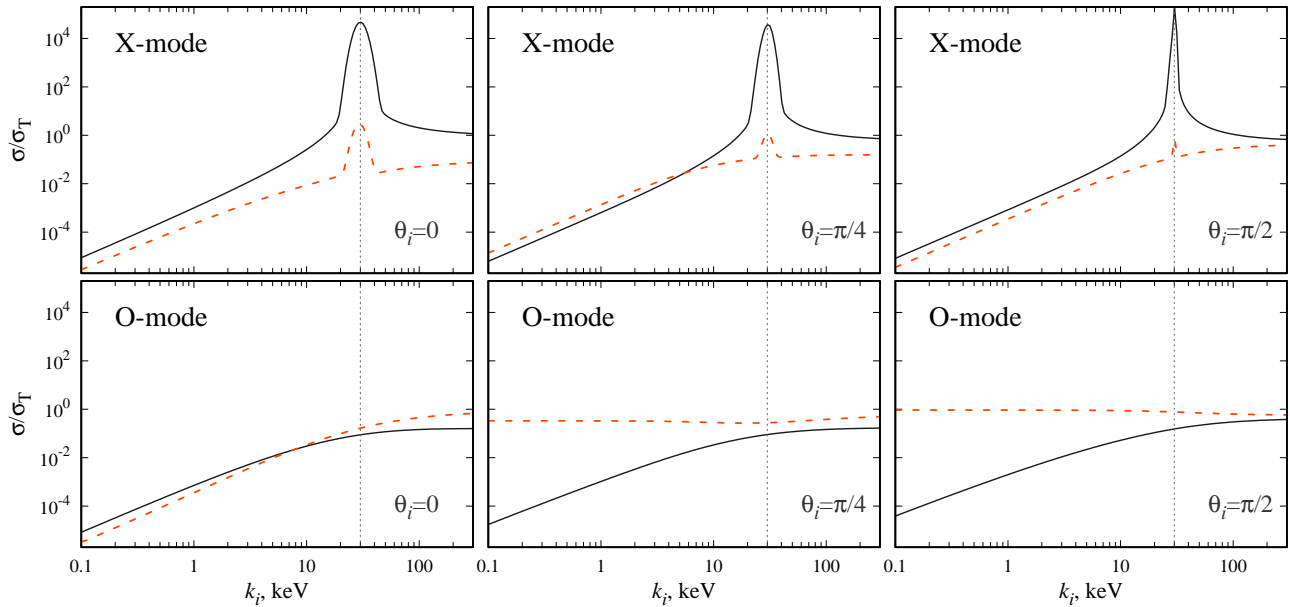} 	
\caption{
Non-relativistic scattering cross section as a function of photon energy at various directions of photon momentum with respect to the magnetic field lines: $\theta_{i}=0$ (left panels), $\theta_{i}=\pi/4$ (central panels) and $\theta_{i}=\pi/2$ (right panels). 
The upper and lower panels show the scattering cross sections calculated for incoming X-mode and O-mode photons, respectively.
The solid black lines correspond to the scattering into X-mode, while the dashed red lines corresponds to the scattering into O-mode.
Parameters: $E_\mathrm{cyc}=30$~keV, $T=5$~keV, $\beta_0=0$.
}
\label{pic:CS_01}
\end{figure*}

If the scattering amplitudes $\widehat{a}$ are known for linearly polarized (vacuum) modes (see expressions for the non-relativistic case in Appendix~\ref{sec:Non-relativistic_apm}), then for arbitrary elliptically polarized modes they can be obtained as follows:
\beq\label{eq:a_v2a_p}
\widehat{a}^{(p)}= \widehat{M}_\mathrm{pv}(E_{f},\theta_{f})\,  \widehat{a}^{(v)}\, \widehat{M}^{-1}_\mathrm{pv}(E_{i},\theta_{i}),
\eeq
where the superscripts $(v)$ and $(p)$ denote the vacuum and plasma polarization cases, respectively, and $\widehat{M}_\mathrm{pv}$ is the transformation matrix given by expressions (\ref{eq:Transform_Matrix}).
The examples of total cross sections 
\beq\label{eq:sigma_int} 
\sigma_{s_\ff s_\ii}(p_{z,i},k_\ii,\theta_\ii)=
\int\limits_{(4\pi)}\d\mathbf{\Omega}_\ff \,
\frac{\d\sigma_{s_{f}s_{i}}}{\d\mathbf{\Omega}_{f}}(p_{z,i},k_{i},\mathbf{\Omega}_\ii,\mathbf{\Omega}_\ff)
\eeq 
between different polarization states for ellipticity given by Eq.(\ref{eq:NormWavEll}) are shown in Fig.\,{\ref{pic:CS_01}}.

\section{Characteristics of redistribution
of photon momenta}
\label{sec:RedistribMomenta}

The redistribution of X-ray photons over the energy, momenta, and polarization states is fully described by the double differential cross section (\ref{eq:DDiffCS00}).
In this paper, we consider three characteristics of the photon redistribution:
\begin{enumerate}{
\setlength{\itemsep}{0pt}
\item the average photon energy $\langle k_{f} \rangle$, 
\item the average photon momentum along $B$-field lines $\langle k_{f}\cos\theta_{f}\rangle$, 
and 
\item the dispersion of the photon energy $D(k_{f})$. 
}
\end{enumerate}
The average energy of a photon after scattering events and dispersion measure of the final photon energy illustrate the features of the Comptonization process and are related to the transformations of the photon energy spectra due to interaction with the electron gas.
The average photon momentum after the scattering events gives an idea about momentum exchange between radiation and gas, which is related to calculations of radiation pressure in a strong magnetic field.

{The total cross section can be  given by
\begin{align}
\label{eq:total}
 &\sigma_{s_{i}}([f_{e}(p_z)],k_{i},\mathbf{\Omega}_{i}) 
 = \nonumber\\
& \sum\limits_{s_{f}}
 \int\limits_{(4\pi)}\d\mathbf{\Omega}_{f}
 \int\limits_{0}^{\infty}\d k_{f}\
 \frac{\d\sigma_{s_{f},s_{i}}}{\d\mathbf{\Omega}_{f}\d k_{f}}([f_{e}(p_z)],k_{i},\mathbf{\Omega}_{i},k_{f},\mathbf{\Omega}_{f}) 
 \nonumber \\
& = \sum\limits_{s_{f}}
 \int\limits_{(4\pi)}\d\mathbf{\Omega}_{f}
 \int\limits_{-\infty}^{+\infty}\d p_z\,f_{e}(p_z)
\frac{\d\sigma_{s_{f},s _{i}}}{\d\mathbf{\Omega}_{f}} (p_z,k_{i},\mathbf{\Omega}_{i},\mathbf{\Omega}_{f}),
\end{align}
where in the second equality we use  Eqs.~(\ref{eq:DiffCrossSection02})
and (\ref{eq:DiffCrossSection03}). }
The averaged energy, longitudinal momentum of a photon, and the dispersion of photon energy after a single scattering are given by
{
\begin{align}
\label{eq:Momenta}
&\left(\begin {array}{ccc}\langle k_{f}\rangle_{1}\\ \langle k_{f}\cos\theta_{f}\rangle_{1} \\D_{1}(k_{f}) \end {array} \right)
= 
\frac{1}{
\sigma_{s_{i}}([f_{e}(p_z)],k_{i},\mathbf{\Omega}_{i})} \nonumber \\
&\qquad\times\left[
\sum\limits_{s_{f}}
\int\limits_{(4\pi)}\d\mathbf{\Omega}_{f}
\int\limits_{-\infty}^{\infty}\d p_{z}\,\,
\left(\begin {array}{ccc}k_{f}\\k_{f}\cos\theta_{f}\\ (\langle k_{f} \rangle-k_{f})^2\end {array} \right) \right. \nonumber \\
&\qquad\qquad
\times \left. 
f_{e}(p_z)
\frac{\d\sigma_{s_{f},s _{i}}}{\d\mathbf{\Omega}_{f}} (p_z,k_{i},\mathbf{\Omega}_{i},\mathbf{\Omega}_{f}).
\right].
\end{align}}
Calculation of the momenta after multiple scattering involves integrals of higher multiplicity, which results in numerical complications.
To overcome them, we use Monte Carlo simulations (see Section \ref{sec:MonteCarlo}), which naturally allow for the multiple scattering and 
cross section modifications due to the photon redistribution.

Since the scattering cross section strongly depends on the photon energy, momentum, and polarization state, the mean free path of photons and the typical time between the scattering events can greatly vary from one photon to another, 
and a typical scattering rate strongly depends on the photon properties. 
We will consider the averaged quantities, affected by the multiple scattering in a given time interval.
We will measure the time intervals in units of a typical time between scatterings, which can be defined, e.g. by the Thomson scattering cross section $\sigma_{T}$, that is $t_{T} = (n_e\sigma_{T})^{-1}$, where $n_e$ is the electron number density.

\begin{figure}
\centering 
\includegraphics[width=1.05\columnwidth]{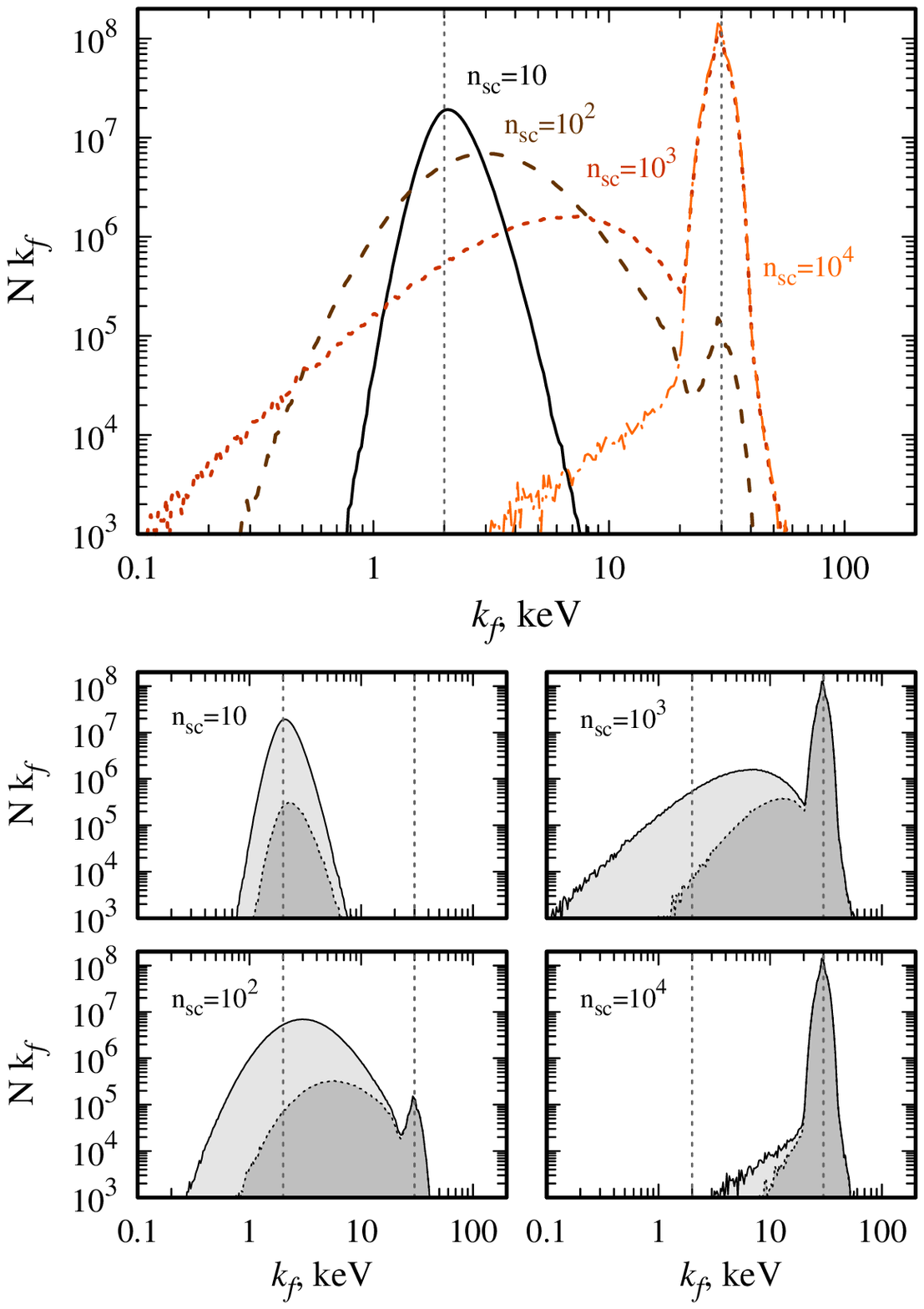} 	
\caption{
Evolution of photon distribution over the energy due to multiple scattering in a strong magnetic field.
The initial photons are taken to be of ellipticity $\xi_1$ (see Eq.\,\ref{eq:NormWavEll}) and propagating initially along the field lines ($\theta_i=0$).
Their distribution over the energy is given by the $\delta$-function at $k_{i}=2$~keV.
Different curves at the top panel describe photon distribution after 
$n_\mathrm{sc}=10$ (the solid line), 
$10^2$ (the dashed line), 
$10^3$ (the dotted line)
and $10^4$ (the dashed-dotted line) scattering 
events. 
The four lower panels represent the same distributions (solid curves) and the parts of the distributions corresponding to photons polarized in the X-mode after $n_\mathrm{sc}$ scatterings (dotted curves and dark grey area below them).
One can see that the photons tend to concentrate within the Doppler core of a cyclotron line and predominantly in the X-mode due to multiple scatterings.
Parameters: $E_\mathrm{cyc}=30$~keV, $T=5$~keV, $\beta_0=0$.
}
\label{pic:sc_spec_evolution}
\end{figure}

\begin{figure*}
\centering 
\includegraphics[width=.9\textwidth]{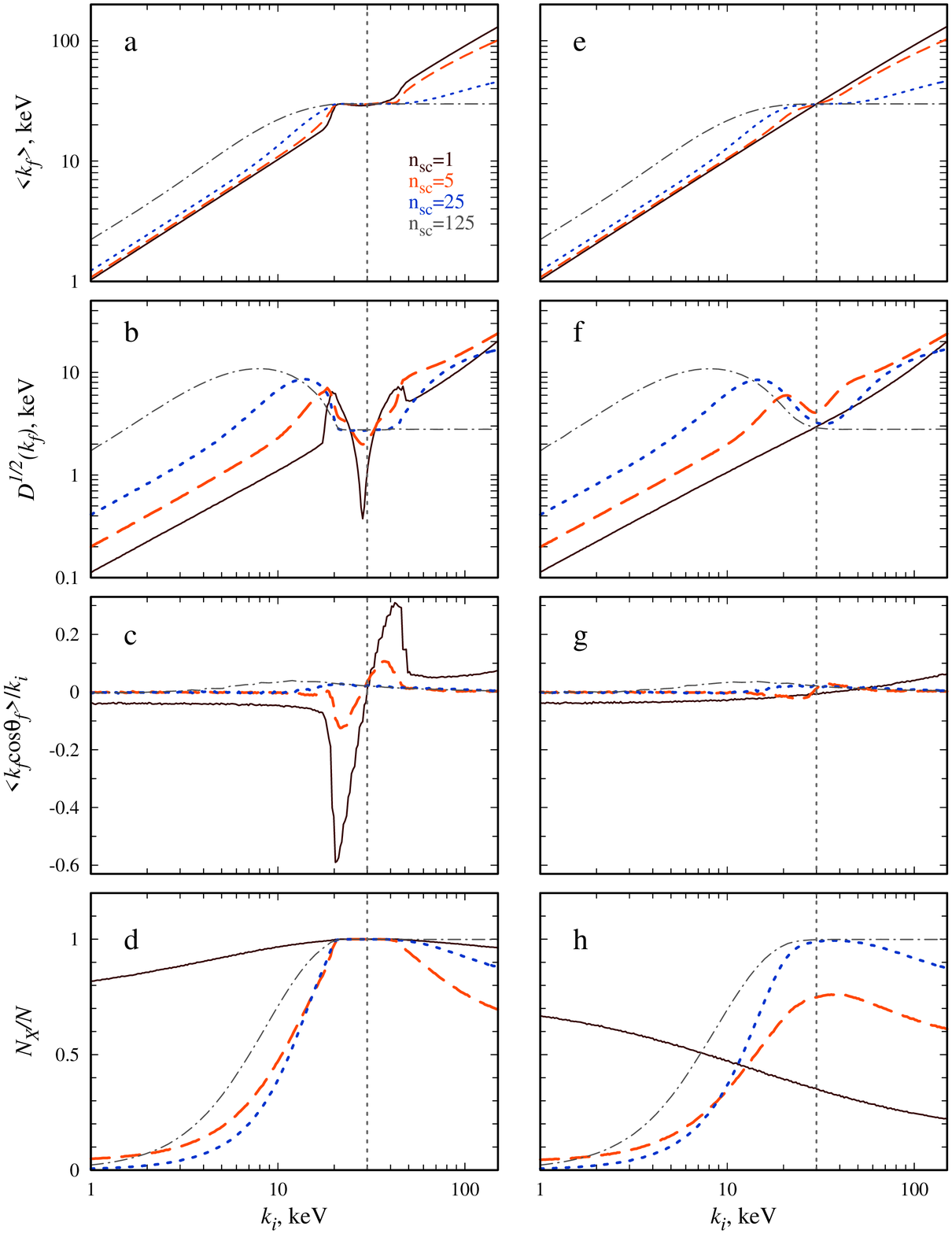} 	
\caption{
The averaged energy of a photon (a,e), the standard deviation of a photon energy (b,f), 
the average photon momentum along the magnetic field (c,g), and the typical fraction of photons polarized in the X-mode (d,h) after
$n_\mathrm{sc}=1$ (black solid lines), 5 (red long-dashed lines), 25 (blue short-dashed lines), and 
125 (grey dashed-dotted lines) scattering events. 
Figures on the left (a,b,c,d) and on the right (e,f,g,h) are given for the photons initially polarized in plasma X- and O-modes respectively. 
The electron gas is taken to be at rest (i.e., $\beta_0=0$).
Parameters: $E_\mathrm{cyc}=30$~keV, $T=5$~keV, $\theta_{i}=0$. 
}
\label{pic:01}
\end{figure*}

\begin{figure}
\centering 
\includegraphics[width=\columnwidth]{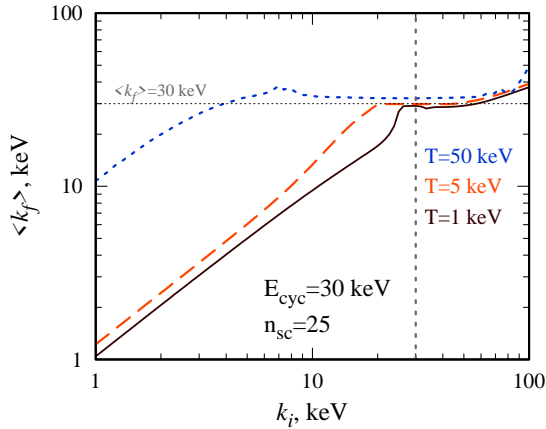} 	
\caption{
The average photon energy after $25$ scattering events by the electron gas of different temperatures:
$T=1$ (the black solid line), $5$ (the red long-dashed line) and $50$~keV (the blue short-dashed line).
One can see that the photons scattered within the Doppler core of a cyclotron line are tending to have the final energy close to the cyclotron one. 
Parameters: $E_\mathrm{cyc}=30$~keV, $\theta_{i}=0$, $\beta_0=0$. 
The run includes $10^6$ photons.
}
\label{pic:Ef_ave_temp_infl}
\end{figure}

\begin{figure*}
\centering 
\includegraphics[width=.9\textwidth]{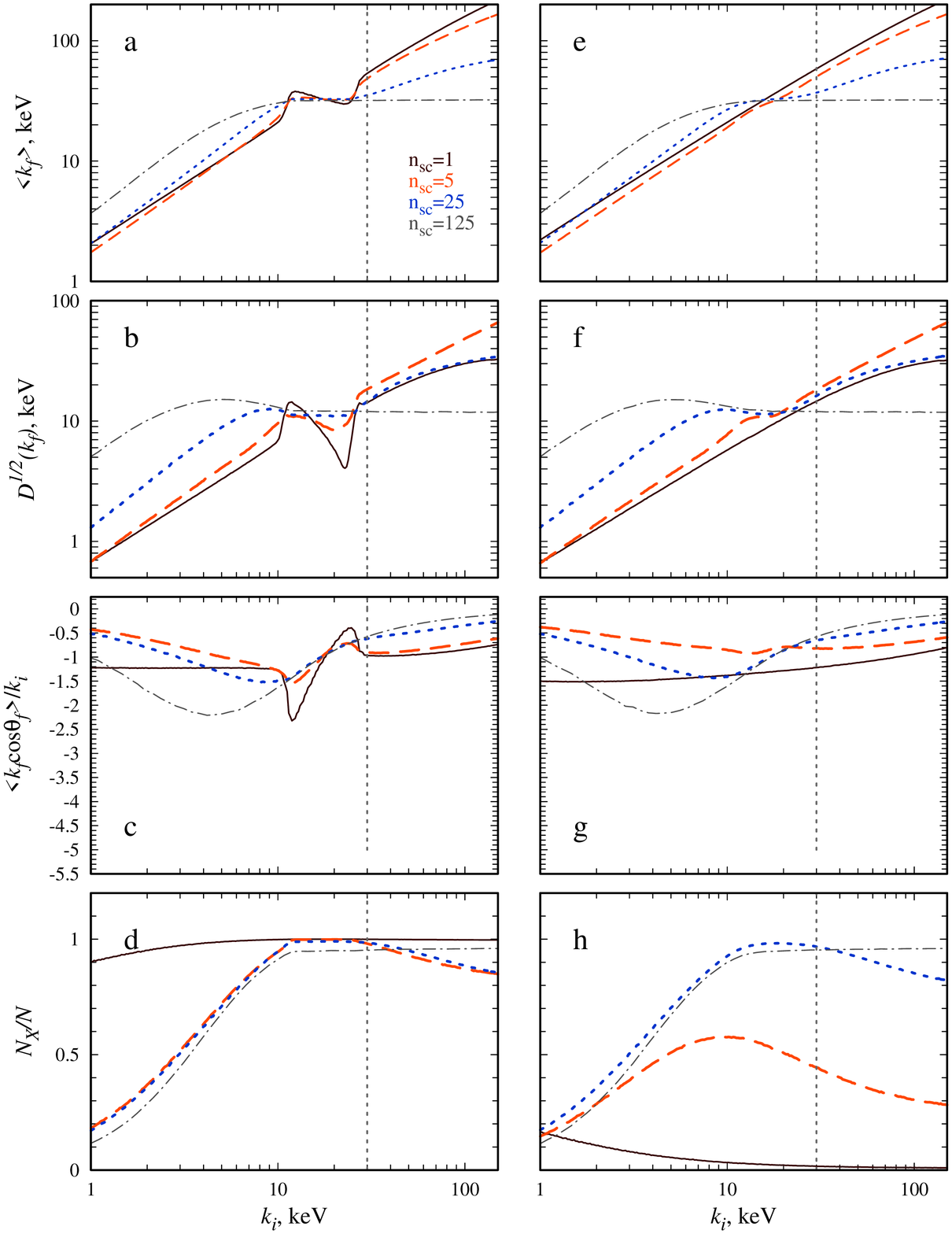} 	
\caption{
The same as in Fig.\,\ref{pic:01}, but the electron gas is taken to be moving with $\beta_0=-0.5$.
}
\label{pic:01_}
\end{figure*}

\begin{figure}
\centering 
\includegraphics[width=\columnwidth]{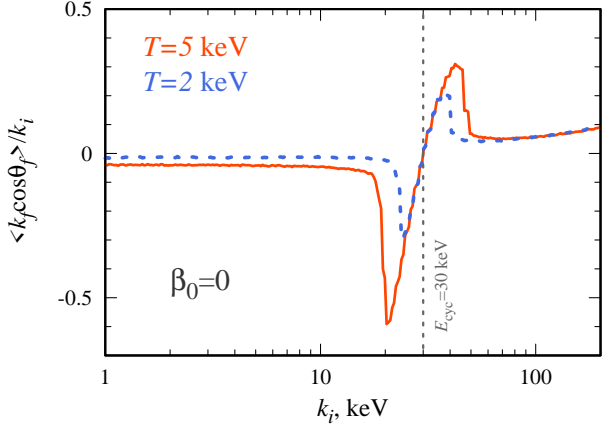} 	
\caption{
The average momentum of a photon along $B$-field lines after a single scattering event.
Different curves are given for different temperatures of electrons: $T=5$~keV (the red solid line) 
and $T=2$~keV (the blue dashed line).
The temperature affects the average momentum after the scattering because the scattering cross section in the laboratory reference frame depends on electron velocity and photons tend to be scattered by electrons with larger cross section.  
Parameters: $E_\mathrm{cyc}=30$~keV, $\theta_{i}=0$, $\beta_0=0$. 
The run includes $10^6$ photons.
}
\label{pic:momentum_on_T}
\end{figure}


\begin{figure}
\centering 
\includegraphics[width=\columnwidth]{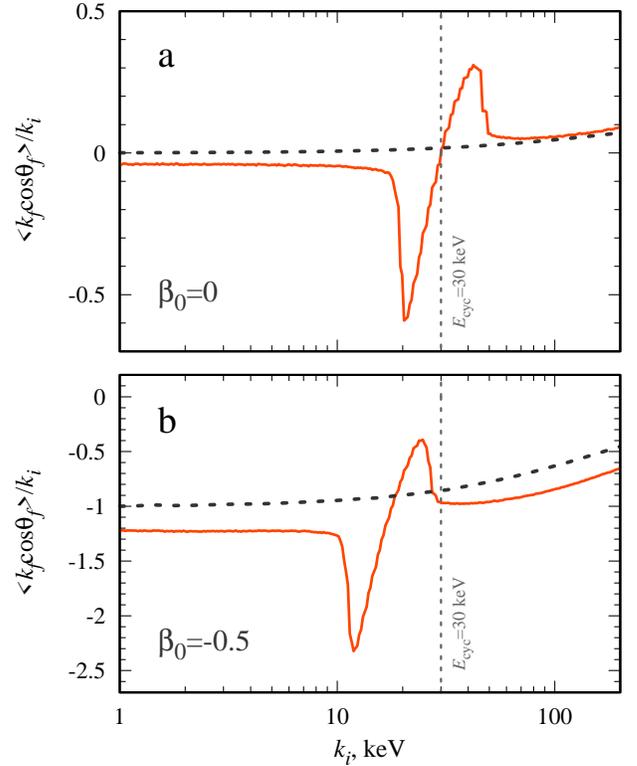} 	
\caption{
The average momentum of a photon along $B$-field after a single scattering (a) in an electron gas at rest or (b) moving with $\beta_0=-0.5$.
Red solid lines represent the scattering of X-mode photons, while black dashed lines represent the scattering in non-magnetic case. 
Parameters: $E_\mathrm{cyc}=30$~keV, $\theta_{i}=0$, $T=5$~keV. 
The run includes $10^6$ photons.
    }
\label{pic:comp_to_non-mag}
\end{figure}

\begin{figure}
\centering 
\includegraphics[width=1.05\columnwidth]{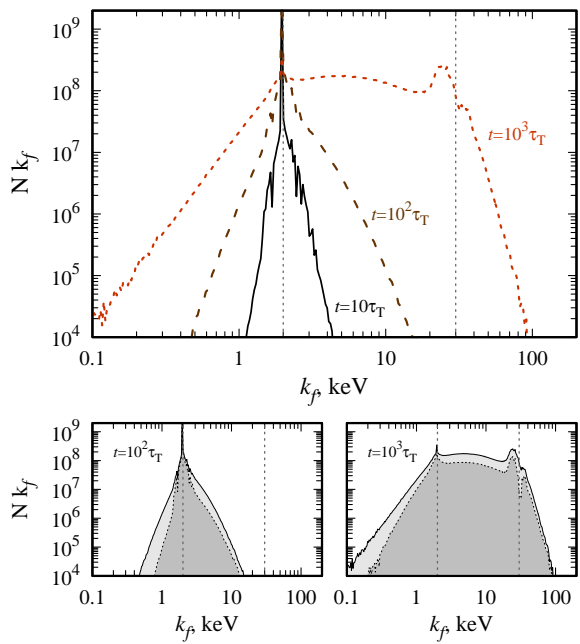} 	
\caption{
Temporal evolution of photon distribution over the energy due to multiple scatterings in a strong $B$-field.
The initial photons are taken to be of pure X-mode propagating along the field lines ($\theta_i=0$). 
Their distribution over the energy is given by the $\delta$-function at $k_{i}=2$~keV.
Different curves at the top panel describe photon distribution after $10\,t_{T}$ (solid), 
$10^2\,t_{T}$ (dashed), and
$10^3\,t_{T}$ (dotted). 
Two bottom panels represent the same distributions (solid curves) and the 
parts of the distributions corresponding to photons of X-mode (dotted curves and dark grey area below them).
Parameters: 
$E_\mathrm{cyc}=30$~keV, $T=5$~keV, $\beta_0=0$.
}
\label{pic:sc_spec_evolution_t}
\end{figure}

\begin{figure*}
\centering 
\includegraphics[width=.9\textwidth]{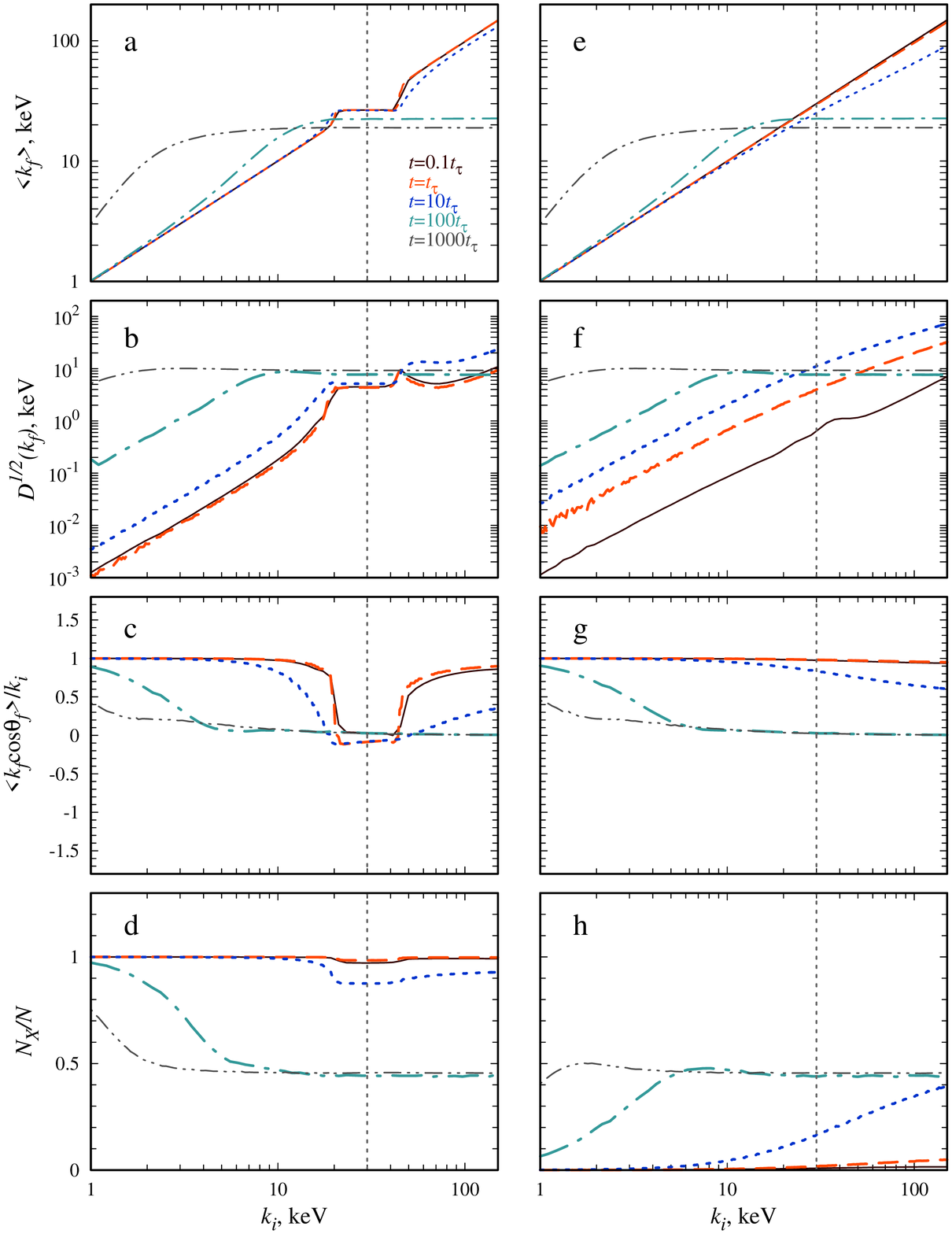} 	
\caption{
The averaged energy of a photon (a,e), the standard deviation of a photon energy (b,f), the averaged photon momentum along magnetic field (c,g), 
and the typical fraction of photons polarized in the X-mode (d,h) after 
{$0.1t_T$} (black solid lines),\,
{$1t_T$} (red dashed lines),\,
{$10t_T$} (blue short-dashed lines),\, 
{$100t_T$} (cyan dashed-dotted lines)\,  
and $1000t_{T}$ (black dashed-double-dotted lines), 
where $t_{T}$ is a typical time scale between two Thomson scatterings.
Figures on the left (a,b,c,d) and on the right (e,f,g,h) are given for the photons initially polarized in plasma X- and O-modes respectively.
Parameters: $E_\mathrm{cyc}=30$~keV, $T=5$~keV, $\theta_{i}=0$, $\beta_0=0$.
The run includes $10^6$ photons.}
\label{pic:02}
\end{figure*}

\begin{figure*}
\centering 
\includegraphics[width=.9\textwidth]{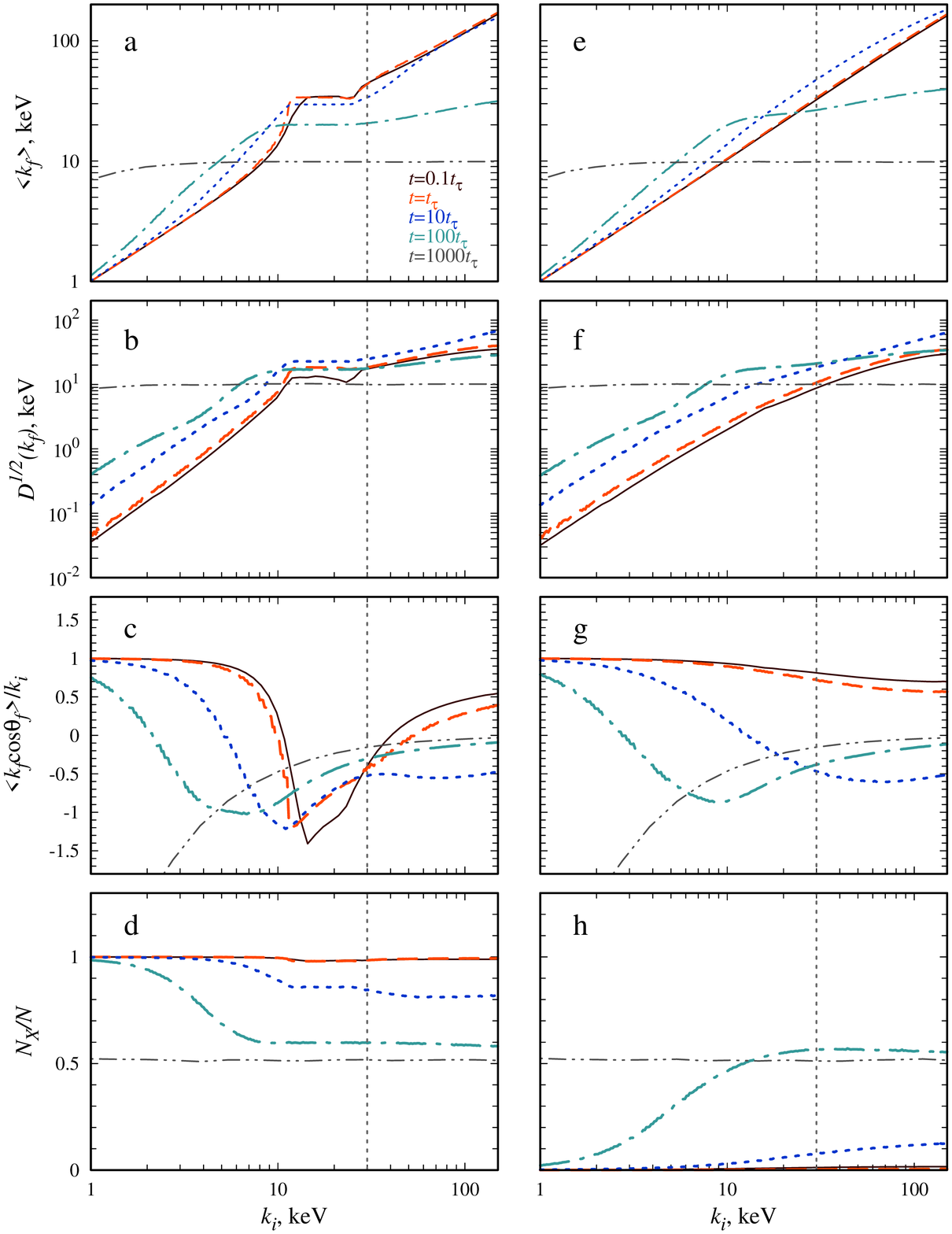} 	
\caption{
The same as in Fig.\,\ref{pic:02}, but the electron gas is taken to be moving with $\beta_0=-0.5$ .
}
\label{pic:02_}
\end{figure*}

\section{Monte Carlo code}
\label{sec:MonteCarlo}

Our Monte Carlo simulations are based on non-relativistic scattering cross sections (see Appendix~\ref{sec:Non-relativistic_apm}) calculated for two polarization modes of a given ellipticity (\ref{eq:NormWavEll}), which is
taken to be determined by the plasma only (see Section \ref{sec:PolModes}). 
We assume that the electrons occupy only the ground Landau level and their distribution over $p_z$, Eq.~(\ref{eq:Maxwell_moving_electrons}), 
is determined by temperature $T$ and bulk velocity $\beta_0$ (see Appendix\,\ref{App:Maxwell}).
The photon redistribution over the momentum, energy and polarization states is affected by the properties 
of electron gas (its temperature and bulk velocity) and initial conditions of a photon.

We perform Monte Carlo simulations and track the photons using a set of the following pre-calculated tables.

(i) \emph{Tables A} give the total scattering cross section as a function of photon energy $k$,
polarization state (X or O)
before  and after  the scattering event, the angle $\theta_i$
between the field and the initial
photon momentum, temperature $T$, and bulk velocity $\beta_e$ of the electron gas.
For each combination of the initial and final polarization states of a photon, the tables are pre-calculated 
in a fixed grid in $k$ and $\theta_{i}$, and for fixed values of $T$ and $\beta_0$.
To get scattering cross section for a given $\bm{k}_i$ we use quadratic interpolations {first} in the photon energy grid and then in the angle grid.

(ii) Extended \emph{Tables B} give probabilities for a photon of a given initial and final polarisation state to be scattered from a given initial direction $\theta_i$ into a certain segment of the solid angle $(\theta_{f}+\Delta\theta_{f},\varphi_{f}+\Delta\varphi_{f})$.
The tables are pre-calculated on a grid of photon initial parameters (energy, momentum, and polarization state) 
and for both possible final polarization states. 

Using \emph{Tables A}, we estimate the total scattering cross section $\sigma$ for a given set of 
parameters and get the mean free path of a photon in units of the Thomson-scattering mean free path, 
$\ell_{T} = (n_e\sigma_{T})^{-1}$.
The actual free path of a photon is taken to be 
\beq
\ell \propto -\ln \eta_1.
\eeq
Here and hereafter, 
$\eta_i\in (0;1)$ is a set of random numbers generated in the Monte Carlo simulations.
In the case of known initial coordinates of X-ray photon, we also get location of the scattering event.
Using the total scattering cross section of a photon of a given physical conditions into X-mode: 
$\sigma_{1 s_\ii}(k_\ii,\theta_\ii)$,
and into O-mode:
$\sigma_{2 s_\ii}(k_\ii,\theta_\ii)$, and generating a random number $\eta_2$, 
we get the photon polarization state after the scattering event: the photon is scattered into X-mode in the case of 
\beq 
\eta_2<\frac{\sigma_{1 s_\ii}(k_\ii,\theta_\ii)}{ \sigma_{1 s_\ii}(k_\ii,\theta_\ii) + \sigma_{2 s_\ii}(k_\ii,\theta_\ii) }
\eeq
and into O-mode in the opposite case. 
At this step, we already have the location of the scattering 
event and photon polarisation state after the scattering.

Using \emph{Tables B} and known initial and final polarisation states we 
generate a random number $\eta_3$ and get the direction of a photon after the scattering event.
Because the electrons are distributed according to $f_e(p_z)$, the direction of the photon momentum 
after a scattering does not determine the final photon energy unambiguously.
In order to obtain the final photon energy, we generate a random number $\eta_4$ and using 
Newton's method get the electron momentum $p_{z,i}$ before the scattering event:
\beq 
\eta_4 = \frac{
\int\limits_{-\infty}^{p_{z,i}}\d p_{z}\,f_{e}(p_{z})
\frac{\d\sigma_{s_{f},s_{i}}}{\d\mathbf{\Omega}_{f}} (p_{z},k_{i},\mathbf{\Omega}_{i},\mathbf{\Omega}_{f})}
{\frac{\d\sigma_{s_{f},s_{i}}}{\d\mathbf{\Omega}_{f}} ([f_{e}(p_z)],k_{i},\mathbf{\Omega}_{i},\mathbf{\Omega}_{f})}.
\eeq
Based on the obtained electron momentum before the scattering event and using the conservation laws (\ref{eq:ConsLaws}), we get the final photon energy.

As a result, we arrive to the final momentum, energy, and polarization state of a photon, using the pre-calculated tables generating four random numbers.
The described scheme allows us to track either photons undergoing a fixed number of scattering events, or photons participating in scattering events during a given time interval, or photons passing a certain distance in a scattering medium.
The scheme can be easily implemented in the models of atmospheres if one applies appropriate boundary conditions (see, for example, Ref.~\cite{2021MNRAS.503.5193M}).
Applying repeatedly the described numerical scheme to the photon with fixed initial parameters, we get different final parameters of the photon and are able to investigate statistical features of multiple Compton scattering.
The examples of spectra and polarization evolution due to the multiple scattering events are presented in Fig.\,\ref{pic:sc_spec_evolution}.

\section{Numerical results}
\label{sec:NumRes}

In this section we describe and discuss the results of our numerical simulations of the statistical features of multiple Compton scattering in a strong magnetic field.
We investigate how the photon distributions over energy, momenta, and polarization state evolves due to the multiple scatterings if the initial distribution is given by the delta function.
The example of such an evolution is represented on the upper panel of Fig.\,\ref{pic:sc_spec_evolution}, where the initial photons are taken to be of equal energy $E=2\,{\rm keV}$.
One can see that the scatterings result in a gradual widening of the distribution over the energy range. 
After a sufficiently large number of scatterings, $n_\mathrm{sc}$, the photons get into the Doppler core of a cyclotron line and stay there.
The four lower panels of Fig.\,\ref{pic:sc_spec_evolution} present the analogous distributions, but resolved into two polarization modes.
It can be seen that inside the Doppler
core of the cyclotron line the X-mode dominates, while
outside the core at lower energies the O-mode 
has a significant advantage.

On the base of the simulated evolution of the photon distributions, we calculate the basic statistical characteristics: the average photon energy, the standard deviation of the photon energy, the average photon momentum along the $B$-field lines, and the average polarization state after multiple scattering events 
({they are shown in Figs.~\ref{pic:01}, \ref{pic:01_}, \ref{pic:02}, and~\ref{pic:02_} below}).
Let us note for clarity that, in contrast with Figs.~\ref{pic:sc_spec_evolution} and \ref{pic:sc_spec_evolution_t}, in all other figures below just  the initial photon energy $k_i$ is plotted along the abscissa axis. 

Starting with the analysis of statistical features of photon distribution after a fixed number of scattering events, we consider later the time evolution of the statistical features (see Section\,\ref{sec:momenta_after_time_int}).
Many specific features of magnetic scattering arise after the first scattering event already. 
The analysis of the time evolution is not straightforward and requires accounting for the dependence of photon free path on the photon momentum and polarisation, which vary from one scattering event to the another.
Meanwhile, the temporal evolution of the statistical features has clear physical meaning. 

The momenta of photon redistribution function 
have been evaluated using
$\sim 10^6$ photons for each set of initial parameters.

\subsection{The energy of X-ray photons after multiple scattering events}
\label{sec:ave_E}

The electron gas at rest scatters the photons below the cyclotron energy with relatively insignificant changes of their energy (see Fig.\,\ref{pic:01}a). 
However, the dispersion of final photon energy increases with the increase of the number of experienced scattering events (see the upper panel in Fig.\,\ref{pic:sc_spec_evolution} and Fig.\,\ref{pic:01}b).
The photons below the cyclotron energy tend to be scattered at large angles regarding the $B$-field direction. 
Because of that, the average momentum of the scattered photons along the field direction is small (Fig.\,\ref{pic:01}c).

The photons having energies within the Doppler core of the cyclotron line take part in resonant scatterings.
The resonant scattering within the Doppler core of the line
effectively produces photons of energy equal to the cyclotron energy (see Fig.\,\ref{pic:sc_spec_evolution}, where the photons tend to stay within the Doppler core as soon as they get there), thus
shifting the averaged energy to the cyclotron one 
(see {Figs.\,\ref{pic:01}a and~}\ref{pic:Ef_ave_temp_infl}).
This happens because the photons ``choose'' the electrons from the available distribution, which provide a larger scattering cross section 
(see expression inside the integral in Eq.\,\ref{eq:DiffCrossSection02}), i.e., the probability of the scattering of the  photon with an energy  $k_i$ and polarization $s_i$ by an electron of momentum $p_{z,i}$ is
\begin{align}
\label{eq:probability_p_z}
&P_{s_i} (p_{z,i},k_{i},\mathbf{\Omega}_{i})
\propto f_{e}(p_{z,i})
\sum\limits_{s_{f}=1}^{2}\sigma_{s_{f},s_{i}} (p_{z,i},k_{i},\theta_{ i}).
\end{align} 
Thus, the photons within the Doppler core of the cyclotron line
are scattered with higher probability by faster oncoming electrons, if the photon energy is smaller than the cyclotron one, $k<E_\mathrm{cyc}$. Vice versa, the photons having $k>E_\mathrm{cyc}$ within the Doppler core are more likely scattered by outgoing electrons.
The standard deviation of the final photon energy varies significantly within the Doppler core (see Fig.\,\ref{pic:01}b),
in contrast to the model of complete redistribution \cite{2019Ap.....62..129G}, which implies that both the average photon energy and the standard deviations are constant within the line core (see, e.g., \cite{1973trsl.book.....I,1978stat.book.....M}).

Photon scattering by the electron gas of non-zero bulk velocity 
shows different features of photon energy redistribution.
In Fig.~\ref{pic:01_} the bulk velocity in the $z$ direction (along the field lines) is chosen  as $\beta_0=-0.5$.
The photons {with energy $E$} below the cyclotron {energy $E_\mathrm{cyc}$}
tend to be scattered with significantly higher
final energies than the initial one 
(see Fig.\,\ref{pic:01_}a 
and compare with Fig.~\ref{pic:01}a ).
This effect causes bulk Comptonization in XRPs \citep{2007ApJ...654..435B}.
The energy of resonant scattering changes due to the Doppler effect, and photons experience resonant scattering at lower energies in the case of their motion oncoming towards the electrons.
This effect is one of the possible causes of variation in cyclotron line centroid energy in sub-critical XRPs \citep{2015MNRAS.454.2714M}.

\subsection{The momentum of a photon along the magnetic field lines}
\label{sec:ave_Pz}

The averaged momentum of a photon after a scattering along $B$-field lines together with the photon momentum before the scattering give a typical momentum which is transferred to an electron due to the scattering:
\beq
\Delta p_{e,z}=k_\ii\cos\theta_\ii-\langle k_{f}\cos\theta_{f}\rangle.
\eeq
A strong magnetic field affects the momentum, energy exchange between particles and the photon redistribution over the directions due to the scattering. 

Because the scattering cross section strongly depends on the photon energy, the average photon momentum after the scattering is affected both by temperature of the electron gas (see Fig.\,\ref{pic:momentum_on_T}) and by bulk velocity (see Fig.\,\ref{pic:comp_to_non-mag}).
The dependence on temperature arises due to the Doppler effect and a strong dependence of the scattering cross section on the photon energy below and near the cyclotron resonance (mainly for the photons of X-mode):
the photons are most likely scattered by those electrons from the distribution, which provide the larger scattering cross section (see Eq.\,\ref{eq:probability_p_z}).
Therefore, resonant electrons are predominantly oncoming for photons with energies below the cyclotron energy, while for photons with higher energy the resonant electrons are predominantly outgoing.
The photons at the red wing of the Doppler core of the resonance tend to get the momentum in the direction opposite to the initial, 
and vice versa, the photons at the blue wing of the core get an extra momentum in the direction of their initial motion (see Figs.~\ref{pic:01}c and ~\ref{pic:momentum_on_T}).  
In the hotter electron gas case, the diversity of electrons of different momenta along the field lines available for scattering is larger.  
Because of that, the momentum exchange between the electron gas and photon gas is more intensive at higher temperatures.
This feature, however, disappears after a few scatterings in the case of the electron gas at rest ($\beta_0=0$). 
The dependence on the bulk velocity is mainly due to the Doppler 
effect, which also determines the photon energy in the reference frame co-moving with the electron.

\subsection{Polarization of X-ray photons under the influence of Compton scattering}
\label{sec:ave_Pol}

Multiple scattering events lead to a gradual decrease of a fraction of X-mode photons and accumulation of O-mode photons below the cyclotron resonance (see the lower panels in Figs.\,\ref{pic:01}, and  \ref{pic:01_}).
This result agrees with rough estimations of the probability of photon polarization transitions.
Indeed, at photon energy well below the cyclotron energy
($E\ll E_\mathrm{cyc}$), the scattering cross section, integrated over angles, can be estimated by order of magnitude as $\sigma^\mathrm{tot}_{O\to O}\sim 0.25\sigma_{T}$ 
for the transition from O-mode into O-mode 
and
$\sigma^\mathrm{tot}_{O\to X}\sim \sigma^\mathrm{tot}_{X\to X}\sim \sigma^\mathrm{tot}_{X\to O}\sim (E/E_\mathrm{cyc})^2\,\sigma_{T}$
for all other transitions (see, e.g., 
\citep{1979PhRvD..19.2868H,1982Ap&SS..86..249K,1995ApJ...448L..29M}). Hence
the probability of photon polarization switch from the O-mode to the X-mode is smaller than the probability to remain in the O-mode 
by a factor $\sim (E/E_\mathrm{cyc})^2$, 
while for the X-mode photons both probabilities to change their polarization mode or to retain it are comparable by order of magnitude.
Therefore, at low photon energies, it is the O-mode that dominates until the number of scatterings becomes so large that gradual 
diffusion 
(due to the exchange of energies between electrons and photons)
of low energy photons 
towards the resonance energy
leads to accumulation of the X-mode (such a tendency can be noticed
in Figs.~\ref{pic:sc_spec_evolution} and \ref{pic:01}). 

Near the cyclotron resonance the situation is opposite.
The X-mode conserves due to the resonant scattering in the vast majority of cases (see the upper panels in Fig.\,\ref{pic:CS_01}), while the scattering of O-mode photons produces the photons of X- and O-modes with 
much lower probabilities (see the lower panels in Fig.\,\ref{pic:CS_01}). 
Thus, multiple scattering 
leads to gradual replacement of the O-mode by the X-mode in a wide vicinity of the cyclotron energy (see Fig.\,\ref{pic:sc_spec_evolution}).

Note that after a sufficient number of scattering events (in our simulations it happens after 25 scatterings already) the 
photon energy distributions become similar for different initial states of photon polarization (compare blue dotted lines from left and right panels on Fig.\,\ref{pic:01},\ref{pic:01_}).

\subsection{Features of photon redistribution after fixed time intervals}
\label{sec:momenta_after_time_int}

The scattering cross section and, therefore, the free path of X-ray photons are strongly dependent on the photon energy, momenta and polarization state. 
All of these parameters are variable due to the scattering events. 
As a result, the temporal evolution of initial photon distribution (see Fig.\,\ref{pic:sc_spec_evolution_t}) is different from the evolution, where the unit step corresponds to the scattering event.

The main feature of photon distributions when considering the multiple photon scatterings in a given time interval is the large difference in the number of scatterings occurring with photons of different energies.
So for relatively long time intervals 
($t=100$ and even $1000~t_{T}$ as in Fig.~\ref{pic:sc_spec_evolution_t})
low-energy photons experience a relatively small number of scatterings until they diffuse (due to the exchange of energies with electrons) in the direction of cyclotron resonance, where the number of scattering increases sharply.
As a result, at $t=10^3~t_{T}$ an approximately uniform distribution is established between the initial photon energy (2~keV in Fig.~\ref{pic:sc_spec_evolution_t}) 
and resonance energy $E_{cyc}$. 
Moreover, a certain balance is maintained between X- and O- modes photons (see below). 

Figs.~\ref{pic:02} and \ref{pic:02_}  (similar to Figs.~\ref{pic:01} and \ref{pic:01_}) show the main 
characteristics of photon redistribution 
in a wide range of initial energies $k_{i}$
after given time intervals.
In general, the characteristics of the photon  redistribution show some features similar to the ones calculated for photons after a fixed number of scattering events.
In particular, the photons within the Doppler core of a cyclotron line still tend to be scattered close to the line center in the reference frame where the electron gas is at rest.
At the same time, strong variations of photon energy deviation from the average value within the Doppler core of a line are reduced (see Figs.\,\ref{pic:02}b,\,\ref{pic:02_}b and compare them with Figs.\,\ref{pic:01}b,\,\ref{pic:01_}b). 
In the limiting case of large time intervals, the 
characteristics of photon redistribution become independent on the initial polarization state of the photon characteristics (compare the values represented on the left and right panels of Fig.\,\ref{pic:02},\ref{pic:02_} corresponding to $t\gtrsim 100\,t_{T}$). 

Futhermore the photon's average polarization state sets at a value different from the one we get at a large number of scattering events (compare Figs.~\ref{pic:02}d and \ref{pic:01}d).
This difference can be explained by the dependence of magnetic Compton scattering cross section on photon polarization state.
After sufficiently large number of scattering events, most of the photons' energies come to the Doppler core of a cyclotron line. 
The scatterings produce photons of the X-mode with a larger probability than photons of the O-mode. 
So, after a large but fixed number of scatterings,
most of the photons are in the X-mode (see Fig.\,\ref{pic:01}d,h).
Still there is a small probability that a scattering will produce photons of the O-mode. Such photons, however,
do not undergo the resonant scattering, and their free path time is larger than that of the X-mode photons.
As a result, the photons of the O-mode are produced rarely but remain in the medium  longer than the photons of the X-mode. 
For this reason, the O-mode photons,
which experienced only a few scattering events,
accumulate in the system during a given  time.
As a result one can notice a tendency to some balance between X- and O-modes in Figs.~\ref{pic:02}  and \ref{pic:02_} for long time intervals.
Note that such an accumulation of O-photons within the medium is a consequence of the fact that the optical thickness is assumed to be infinitely large, otherwise the polarization ratio at
$t > t_T$ would be principally different.

\section{Astrophysical applications}
\label{sec:AstroApp}

The present study aims at description of radiative transfer in isolated and accreting strongly magnetized NSs. 
In this respect, the following points are of particular interest:
\begin{itemize}
{\setlength{\itemsep}{0pt}
\item Radiative transfer in a cyclotron line is strongly affected by thermal and bulk motion of electrons. 
The photon redistribution within the Doppler core of a line is significantly different from the one given by the approximation of complete redistribution.
This result is particularly important for numerical simulations of X-ray spectra and investigation of cyclotron line features in NSs \citep{2019A&A...622A..61S,2013ApJ...777..115P,2014ApJ...781...30N,2015MNRAS.454.2714M,2019PASJ...71...42N,2021arXiv210807573K}. 

\item 
Confinement of X-ray photons within the Doppler core of a cyclotron line can affect Comptonization of soft X-rays by hot electron gas, preventing appearance of high-energy tails with exponential cut-off at $E\sim T$ in the energy spectra.
Note that this statement still has to be tested 
in realistic geometrical settings, taking boundaries of the scattering region into account.
Here we do not consider any boundary effects, hence our present analysis 
of multiple photon scattering is only applicable well inside the scattering region.

\item The efficiency of momentum transfer from the photons to the electrons due to the Compton scattering in a strong magnetic field differs significantly from that in the non-magnetic case. 
A strong dependence of the scattering cross section on
photon energy below the cyclotron resonance results in more efficient momentum transfer to the gas moving towards the source of photons.
The efficiency of momentum transfer depends on the gas temperature both within the Doppler core of the cyclotron line and below the cyclotron resonance, i.e., within the photon energy intervals where the scattering cross section strongly depends on the photon 
momentum. 
The specific effects of momentum transfer come into play after the first scattering event already and, therefore, are valid both for optically thin and optically thick medium.
Such features of momentum transfer from the photons to the gas are particularly important for calculations of radiation pressure in accreting strongly magnetized NSs and for estimations of the critical luminosity sufficient to stop accretion flow above the stellar surface \citep{1976MNRAS.175..395B,2015MNRAS.447.1847M} and structure of the accretion columns at high mass accretion rates in XRPs \citep{1981A&A....93..255W,1990ApJ...349..262B,2015MNRAS.454.2539M,2021MNRAS.504..701B}. 
}

\end{itemize}

\section{Summary}
\label{sec:Summary}

We have considered statistical features of magnetic Compton scattering of polarized X-ray photons by the electron gas,
taking into account the resonant scattering at the fundamental cyclotron frequency, thermal distribution of electrons
on the ground Landau level, and the bulk motion of the electron gas.
We have used an approximation of non-relativistic scattering cross sections and photon dispersion relations dominated 
by plasma effects.
Effectively, multiple scattering events were investigated in the infinite medium, i.e., we did not consider the process of photon's escape from the scattering region.
This analysis allows us to make conclusions on the features of photon redistribution and momentum transfer from photons to the medium due to the interaction with the electron gas inside the scattering region.
Statistical feature of single scattering events can be applicable to the analysis of radiative transfer in optically thin medium, when photons hardly experience more than one scattering.

We have shown that the photons with initial energies $E$ in a wide vicinity of the cyclotron energy $E_\mathrm{cyc}$ tend to acquire final energies closer to $E_\mathrm{cyc}$ after scattering (see Figs.\,\ref{pic:01}a,e, \ref{pic:02}a,e) with relatively small dispersion measure (see Figs.\,\ref{pic:01}b,f,\,\ref{pic:02}b,f). 
It means that the photons can be confined near the cyclotron energy, where the scattering cross section is large.
This confinement of $E$ near $E_\mathrm{cyc}$ increases the probability of true absorption due to the cyclotron mechanism and due to free-free transitions amplified at cyclotron energy in a strong magnetic field (see, e.g., \cite{1992herm.book.....M,2015SSRv..191..171P} and references therein).
Photon redistribution within the Doppler core of a cyclotron line is significantly different from the complete redistribution widely used as approximation in radiative transfer calculations
\cite{1973trsl.book.....I,1978stat.book.....M}.  

The efficiency of photon momentum transfer 
strongly 
depends on energy $E$ near $E_\mathrm{cyc}$ 
(see Figs.\,\ref{pic:01}c,g, \ref{pic:01_}c,g).
The photons with $E$ slightly smaller than $E_\mathrm{cyc}$ tend to be scattered by electrons moving towards them, which results in an efficient momentum transfer from the photons to the electrons.
The photons with $E$ slightly larger than $E_\mathrm{cyc}$ are more efficiently scattered by electrons moving in the same direction with them.
The strong dependence of the scattering cross section on the photon energy in vicinity of the resonance results in a dependence of momentum exchange on the electron temperature.
These specific features of momentum transfer arise after the first scattering event already.

Bulk velocity of the electron gas strongly affects exchange of momenta  between electrons and photons (see Fig.\,\ref{pic:comp_to_non-mag}).
In the case of oncoming motion, photons below the cyclotron resonance acquire a negative impulse and thereby transfer inverse impulse to the electron gas more efficiently than in the non-magnetic case.
The braking of accreting material due to a strogh radiative force is typical for bright X-ray pulsars (see, e.g., \citep{1976MNRAS.175..395B,2015MNRAS.447.1847M,2022arXiv220414185M}).
This feature of magnetic Compton scattering is essential for the braking of accreting material, in particular, in the radiation-dominated shock near the NS surface at sufficiently high mass accretion rates.

Multiple Compton scatterings, limited by a given temporal  (or spatial) interval, tend to reduce the fraction of X-mode photons or to  balance the X- and O-modes.
This effect, however, should be considered consistently with the effect of a larger free path of X-mode photons at energies below the cyclotron energy when analyzing problems of the polarized radiation transfer in highly magnetized atmospheres of NSs.

\section*{Acknowledgements}

The work of AAM was funded by the Netherlands Organization for Scientific Research Veni fellowship.
The work of ADK, and AYP was funded by RFBR according to the research project 19-52-12013.
The work of VFS was supported by the German Research Foundation (DFG) grant WE 1312/53-1.
AAM is also grateful to city library in Bussum for its hospitality.
We are grateful to an anonymous referee for a number of useful comments and suggestions which helped us improve the paper.



\appendix

\section{One-dimensional relativistic Maxwell distribution}
\label{App:Maxwell}

The electrons and positrons in a strong magnetic field are distributed over the Landau levels and move along $B$-field lines.
The distribution over the Landau levels in the case of local thermodynamic equilibrium is described by the Boltzmann law. 
Then the majority of electrons occupy the ground Landau level
in the case of extremely strong magnetic field ($E_\mathrm{cyc}\gg T$).
However, in the conditions characteristic of the atmospheres, magnetospheres, and accretion columns
around strongly magnetized NSs, the typical time scale of radiative de-excitation from upper Landau levels is much shorter than the typical time scale of collisional excitation, whence it follows that the majority of electrons occupy the ground Landau level 
not only at $T\ll E_\mathrm{cyc}$, but
even in the case of temperatures comparable to the cyclotron energy (e.g., Refs.~\cite{1979A&A....78...53B,1992herm.book.....M}; 
see also Ref.~\cite{2007MNRAS.376..793P}).

\begin{figure}
\centering 
\includegraphics[width=\columnwidth]{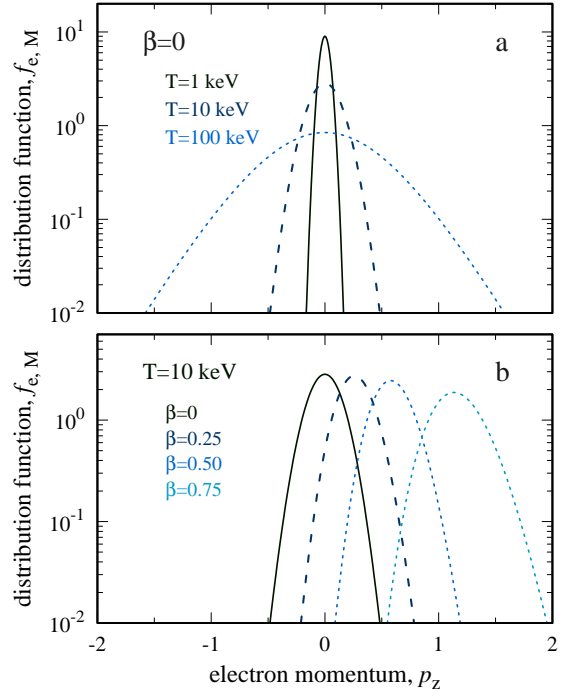} 	
\caption{
The relativistic one-dimensional Maxwellian distribution of electrons over the momentum along magnetic field lines.
Top panel: different curves are given for different temperatures of the electron gas while the bulk velocity is fixed at $\beta_0=0$.
Bottom panel: different curves are given for different bulk velocities of electron gas with temperature fixed at $T=10$~keV.  
}
\label{pic:Maxwell}
\end{figure}

If electrons occupy the ground Landau level only, their one dimensional relativistic Maxwellian distribution in the reference frame co-moving with the gas is given by
\beq 
f_{e}(p_z)=f_\mathrm{e,M}(p_z,T,\beta_0	=0)=\frac{e^{-y\gamma}}{2K_1(y)},
\eeq
where $y\equiv m_e/T$ and $K_{1}(y)$ is the modified Bessel function of the second kind, which provides the normalization
$\int_{-\infty}^{\infty}\d p_z\,f_{e}(p_z)=1$ (see Fig.\,\ref{pic:Maxwell}a).

{The distribution in the arbitrary reference frame can be obtained from the Lorentzian transformation.
The distribution function $f^*_{e}(p_z)$ normalized by the particle number density $N$ is invariant under Lorentz transformation, but the number density is not: we have
\beq
\int_{-\infty}^{\infty}\d p_z\,f^*_{e}(p_z,\beta_0)=N(\beta_0)=\gamma_0 N_0,
\eeq
where $\gamma_0=(1-\beta_0^2)^{-1/2}$ is the gamma-factor due to bulk velocity of an electron gas and $N_0$ is the number density in comoving reference frame.
Therefore, the distribution function normalized by the unity (see Fig.\,\ref{pic:Maxwell}b) is
\begin{eqnarray}
\label{eq:Maxwell_moving_electrons}
f_\mathrm{e,M}(p_z,T,\beta_0)&=&\gamma_0^{-1}f_\mathrm{e,M}(p'_{z},T,\beta_0=0)
\nonumber \\
&=& \frac{e^{-y[\gamma_0(\gamma-\beta_0 p_z)]}}{2\gamma_0 K_1(y)},
\end{eqnarray}
and 
\beq
\int_{-\infty}^{\infty}\d p_z\,f_\mathrm{e,M}(p_z,T,\beta_0)=1.
\eeq
} 

\section{Non-relativistic scattering amplitudes: analytical expressions}
\label{sec:Non-relativistic_apm}

In the case of photon propagation in magnetized vacuum the normal polarization modes are linearly polarized.
The electric vector of the ordinary mode (O-mode, denoted by ``2" here) belongs to the plane composed by the $B$-field direction and direction of photon momentum,
while the electric vector of extraordinary mode (X-mode, denotes by ``1" here) is perpendicular to this plane.

Non-relativistic amplitudes describing the scattering of linearly polarized modes (see e.g. \cite{1979PhRvD..19.2868H}) are given by
\begin{align}
\label{eq:CS_Amp_01}
&a^\mathrm{(v)}_{11}=
g + f, \\
&a^\mathrm{(v)}_{22}= 2S_\ii S_\ff+ C_\ii C_\ff
\left(g+f\right), \\
&a^\mathrm{(v)}_{12}= -i C_\ii \left(g-f\right) , 
\\
\label{eq:CS_Amp_01_}
&a^\mathrm{(v)}_{21}= +i C_\ff \left(g-f\right),
\end{align} 
where 
\[
g=\frac{k_\ii}{k_\ii+E_\mathrm{cyc}}e^{i(\varphi_\ii-\varphi_\ff)},
\quad
f=\frac{k_\ii}{k_\ii-E_\mathrm{cyc}}e^{-i(\varphi_\ii-\varphi_\ff)},
\quad
\]
$C_i= \cos\theta_i$,
$C_f= \cos\theta_f$,
$S_i= \sin\theta_i$,
and
$S_f= \sin\theta_f$. 
The scattering described by the amplitudes (\ref{eq:CS_Amp_01}--\ref{eq:CS_Amp_01_}) is resonant at the cyclotron energy $E_\mathrm{cyc}$, where the denominator $(k_{i}-E_\mathrm{cyc})=0$ and the absolute values of the amplitudes turn to infinity.
These infinities are removed by the regularization procedure \citep{1993A&AT....4..107N}, when one allows for the natural width of Landau levels \citep{1982A&A...115...90H,1991ApJ...380..541P,2005ApJ...630..430B}. 
In our calculations we follow the approximations proposed in \cite{1991ApJ...380..541P} (see Section VI and Appendix B in \cite{2016PhRvD..93j5003M} for more details).
For future use it is convenient to introduce the matrix composed of scattering amplitudes with the following notation:
\beq\label{eq:a_v}
\widehat{a}^\mathrm{(v)}=
\left(\begin {array}{cc} a^\mathrm{(v)}_{11} & a^\mathrm{(v)}_{12} \\ a^\mathrm{(v)}_{21} & a^\mathrm{(v)}_{22} 
\end {array} \right).
\eeq

Using transformation (\ref{eq:a_v2a_p}) one can get the scattering amplitudes for elliptically polarized photons using Eq.~(\ref{eq:a_v}).
The corresponding analytical expressions are
\begin{align}
\label{eq:CS_Amp_02_}
a^\mathrm{(p)}_{11}=&
\varrho \left[ 2\xi_f \xi_i S_f S_i 
+g (1-\xi_i C_i)(1-\xi_f C_f) \right. \nonumber \\
& \left. + f (1+\xi_i C_i)(1+\xi_f C_f)\right], \\
a^\mathrm{(p)}_{22}=& \varrho  \left[2S_i S_f +g (\xi_i+C_i)(\xi_f+C_f) \right. \nonumber \\
& \left. + f (\xi_i-C_i)(\xi_f-C_f) \right], \\
a^\mathrm{(p)}_{12}=& \varrho \left[ 2\xi_f S_f S_i +g(C_i+\xi_i)(\xi_f C_f-1) \right.\nonumber\\ 
&\left. +f(C_i-\xi_i)(\xi_f C_f+1) \right] , \\
a^\mathrm{(p)}_{21}=&\varrho  \left[ 2\xi_i S_f S_i +g(C_f+\xi_f)(\xi_i C_i-1) \right.\nonumber \\ 
&\left.+f(C_f-\xi_f)(\xi_i C_i+1) \right],
\end{align}
where
$
\varrho = (1+|\xi_i|^2)^{-1/2}(1+|\xi_f|^2)^{-1/2}
$
and
$\xi_i=\xi_{X,i}$, $\xi_f=\xi_{X,f}$ are ellipticities of the extraordinary mode before and after the scattering event, respectively.
Note that in a limit of coherent scattering ($k_f=k_i$), only the amplitude $a^\mathrm{(p)}_{11}$ shows resonance at the cyclotron energy, while the other amplitudes are not resonant \citep{1982Ap&SS..86..249K}. 
However, if one takes into account the electron recoil and the corresponding change of photon energy, 
the amplitude $a^\mathrm{(p)}_{21}$ (corresponding to polarization transition from X- to O-mode) also becomes resonant at $k_i=E_{\rm cyc}$.  

\bibliographystyle{apsrev.bst}

\label{lastpage}
\end{document}